# Lidar-based gas analyzer for remote sensing of atmospheric methane

Viacheslav Meshcherinov[a,b]*, Viktor Kazakov[a,b], Maxim Spiridonov[b], Gennady Suvorov[c,d], Alexander Rodin[a]

[a] Moscow Institute of Physics and Technology (National Research University), 141701 Dolgoprudny, Russia
[b] Space Research Institute of the Russian Academy of Sciences (IKI RAS), 117997 Moscow, Russia
[c] Institute of Forest Science of the Russian Academy of Sciences (IFS RAS), 143030 Uspenskoe, Russia
[d] A. N. Severtsov Institute of Ecology and Evolution of the Russian Academy of Sciences (IEE RAS), 119071 Moscow, Russia

*Corresponding author. Email: meshcherinov@phystech.edu

**Abstract**

Enhancement of methane emission measurement techniques is necessary to address the need for greenhouse gas emissions monitoring. Here we introduce a gas analyzer designed for remote sensing of atmospheric methane aboard unmanned aerial vehicles. This device employs the wavelength modulation spectroscopy approach and quadrature detection of laser radiation scattered from the underlying surface. Our results demonstrate that the observed correlation between various signal attributes and the distance to the surface, where laser radiation scatters, aligns with analytical expectations. Calibrations proved that the instrument provides reliable methane measurements up to 120 meters while being lightweight and power efficient. Notably, this device outperforms its competitors at altitudes exceeding 50 meters, which is safer for piloting.

## 1. Introduction

Methane is known to be the second most prevalent anthropogenic greenhouse gas (GHG) following $CO_2$, contributing to 17% of global GHG emissions originating from human activities [1]. The globally averaged monthly mean atmospheric methane abundance data from the U.S. National Oceanic and Atmospheric Administration indicates a ~20% rise in atmospheric methane levels between 1983 and 2023 [2]. This trend is a matter of great concern due to methane's heightened efficacy as a greenhouse gas compared to $CO_2$ – its cumulative GWP exceeds that of $CO_2$ by 28 times over 100 years and by 84 times over 20 years [3,4,5].

Natural methane sources significantly contribute to the escalation of atmospheric methane content. Emissions from natural sources may increase as wetlands warm, tropical precipitation increases, and notably surge with the permafrost thaw. In addition to natural gas, both biogenic and endogenous (geological) sources emit methane into the atmosphere. However, alongside natural pathways of methane introduction into the atmosphere, 50-65% of global $CH_4$ emissions is linked to anthropogenic activities [6].

In the past two decades, the surge in anthropogenic emissions, originating primarily from agricultural and waste utilization activities in South and South-East Asia, South America, and Africa, as well as from fossil fuel usage in China, the Russian Federation, and the United States, has been identified as the predominant driver behind the escalation in atmospheric methane emissions [7]. Reducing methane emissions from the fossil fuel, waste, and agricultural sectors holds the potential for climate change mitigation in the short term. Notably, substantial reductions in $CH_4$ emissions could be attained by addressing leaks in pipelines and industrial facilities associated with the oil and gas production industry [6].

The United Nations Environment Programme's (UNEP) 2021 GHG emissions gap report illustrates that existing national climate commitments, when combined with additional mitigation measures, may result in a global temperature rise of 2.7℃ by the end of the century compared to the beginning of the industrial era [8]. Reducing methane emissions to levels mandated by the Paris Agreement holds promise for significantly attenuating the dynamics of global warming [9].

However, beyond regulatory challenges, the deficiency in robust methods for GHG emissions accounting released into the atmosphere poses a barrier to realizing the objectives outlined in the Paris Agreement. Hence, there is an imperative need to enhance methodologies for measuring GHG emissions, given that current estimation techniques, particularly prevalent in industries such as oil and gas, may underestimate methane emissions into the atmosphere by 1.5-2 times [10].

Identifying substantial methane emission sources, which constitute a significant fraction of emissions from the oil and gas sector, employs a diverse array of approaches. Firstly, there are several satellites already in orbit, such as GHGSat (Canada/UK), GOSAT (Japan) and Sentinel-5P (Europe). Some of them measure $CH_4$ from point sources such as gas leaks, and over localized areas [11,12]. However, in the case of cloud cover of the region of interest, this approach does not work, moreover, the ability of the satellite to detect small methane emission sources needs to be higher [13].

In recent years, methodologies leveraging unmanned aerial vehicles (UAVs) have emerged, offering the advantages of remote sensing without the logistical constraints associated with ground-based monitoring or the limitations of manned aircraft, particularly at altitudes of up to 100 m, which are inaccessible to conventional aircraft, thus enhancing spatial resolution.

Nevertheless, most methane emission monitoring devices mounted on small UAVs are not optimized for remote sensing operations at tens to hundreds of meters at altitudes. These devices encompass *in situ* analysis tools such as contact chemical sensors, multi-sensor arrays capable of detecting various target gasses, and so-called electronic nose (e-nose) systems. Despite their affordability, low power consumption (typically a few mW), and lightweight design (tens of grams), these sensors exhibit limitations including slow response times, modest sensitivity, and susceptibility to cross-sensitivity with non-target gasses [14]. Conversely, certain devices analyze gas mixtures within an analytical cell – examples include laser spectrometers based on cavity ring-down spectroscopy (CRDS) and off-axis integrated cavity output spectroscopy (OA-ICOS). A novel optical scheme suitable for compact *in situ* laser spectrometers, a segmented circular multipass cell (SC-MPC), has been recently demonstrated that allows efficient and interference-free laser beam folding featuring up to 10 m optical path length [15]. Notably, a lightweight iteration of an open CRDS-cell device with low power consumption has recently been developed, suitable for integration onto UAV platforms as well as SC-MPC mid-IR laser spectrometer for airborne, *in situ* atmospheric methane measurements and photo-acoustic gas analyzer, based on quantum cascade laser [16,17,18].

Such gas analyzers are effective in vertical profiling of the concentration distribution of various gas species [19]. But in the case of such applications for natural gas leak detection from pipelines, their fundamental disadvantage is low accuracy of leak localization and high probability of false response. In addition, using this type of leak detection device is often difficult because of the need to consider wind direction and velocity to ensure that the UAV can only move leeward of the leak. At the same time, a safe flight altitude of UAV in urbanized terrain should be considered at least 30-50 m, to avoid collision with trees, power line masts, and other tall objects, as well as for the possibility of making a maneuver in an emergency to save the UAV and payload.

Optical lidar-type gas analyzers founded on absorption spectroscopy present a fitting solution for remote sensing of methane abundance in atmospheric air. Devices employing the principle of receiving radiation scattered from the surface typically position both the emitter and detector on the same side of the optical path. The emission of laser radiation towards a distant surface is accompanied by the collection of scattered light by the photodiode. Diverging from the instruments mentioned above, these devices measure integral gas concentration in units of ppm·m along the laser beam.

Among many UAV-mounted gas analyzers, prominent examples include those produced by Gastar Co. Ltd (Japan), notably the Laser Methane mini (LMm) detector. Originally developed in 1992 by Tokyo Gas Engineering Solutions (Japan) as a portable handheld detector for ground inspections [20], the LMm detector was commercially introduced in 2013 [21] and has since been integrated into various UAV-based remote sensing systems. Tokyo Gas Engineering Solutions has recently introduced the Laser Falcon detector, tailored specifically for UAV applications [22]. These lightweight instruments, weighing up to 1 kg, offer relatively high sampling rates of ≥2 Hz and exhibit accuracy within the range of 100-1000 ppm·m, albeit necessitating low operating altitudes of up to 30 m and reduced UAV horizontal speeds to ensure stable operation [23,24].

Conversely, larger instruments, such as the Pergam Laser Monitoring Fixed Wing (LMF) and Pergam ALMA G4 mini, require either a UAV equipped with a rigid wing or a UAV with a lifting capacity exceeding 10 kg respectively. Despite better methane sensitivity, such massive devices experience limited demand due to higher costs and operational complexity.

In this paper, we introduce the *Gas analyzer for the Investigation of Methane Lidar-based Infrared* (GIMLI), developed by our research group. GIMLI represents a lidar-type spectrometer tailored for remote sensing of integral methane content in atmospheric air from small-capacity UAV platforms.

## 2. Method

The validation of the wavelength modulation spectroscopy (WMS) with quadrature detection (QD) of the received signal method in laboratory settings has been described in our prior publication [25]. QD addresses the challenge of remote measurements of methane abundance in atmospheric air similar to the phase-sensitive lock-in technique, eliminating the phase dependency through the usage of quadratures [26,27]. Subsequently, the project progressed to a prototype instrument based on this methodology development, tailored to meet the specifications for integration onto UAVs with a payload capacity of up to 5 kg. The development and refinement of this instrument spanned approximately three years. Following the completion of the refinement phase, the developed device underwent calibration using data obtained from the LI-COR LI-7810 mobile gas analyzer. Figure 1 illustrates the assembled prototype of the gas analyzer.

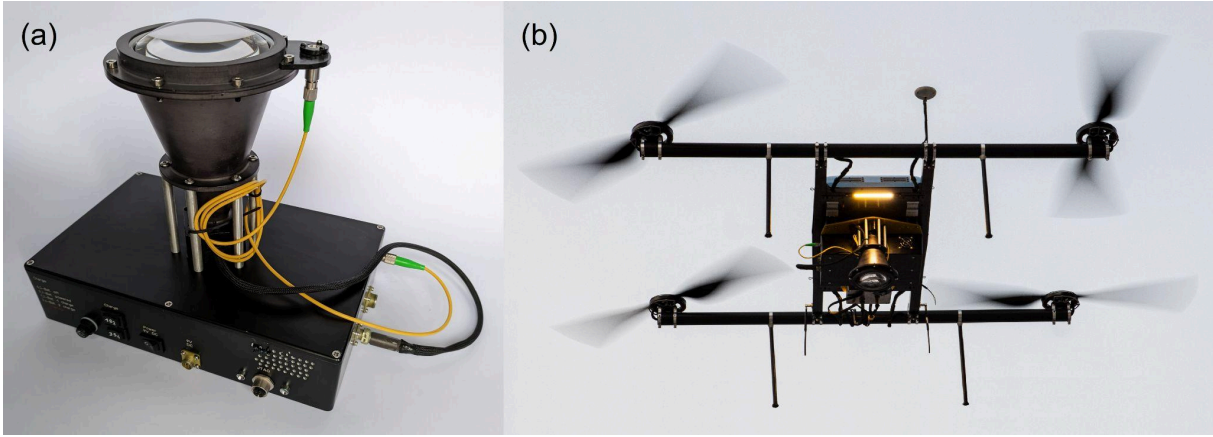

*Fig. 1. (a) Standalone GIMLI prototype. (b) GIMLI prototype aboard the Irbis-432 UAV.*

Given that the standard movement speed of a multirotor UAV typically ranges from 15 to 20 m/s, and the spatial resolution of sensing should match or exceed the characteristic dimensions of the leak in the vicinity of the leakage site – typically measured in units of meters – a methodology capable of achieving a repetition rate of a full cycle of spectral measurements at the level of tens of Hz is essential to address this challenge. Moreover, such a methodology must offer sensitivity adequate to detect deviations from the natural background methane levels by units to tens of percent. In a sparsely urbanized area, the typical methane abundance in ambient air is approximately 1.8 to 2.2 ppm. The altitude of the UAV above the ground surface is typically set at 50 m, thereby resulting in an integral concentration of around 200 ppm·m. Based on the WMS and QD of radiation scattered from the surface application, the proposed technique satisfies these requirements, as demonstrated in [25].

### 2.1. Spectral range

The methane absorption lines near 3.24 μm (3086 cm$^{-1}$), corresponding to the fundamental mode ν3 of C-H bond vibrations, exhibit approximately 40 times greater intensity compared to the overtone lines of this mode near 1.65 μm (6057 cm$^{-1}$). However, this discrepancy is offset by the feasibility of inexpensive fiber optics, compact semiconductor photodiodes, and diode lasers in the near-infrared range. A specific detectivity $D^*$ of such photodiodes in their operational range is three orders of magnitude higher than that of uncooled InAs photodiodes operating in the 3 μm region, and two orders of magnitude higher than that of uncooled HgCdTe photodiodes designed for operation in the 2-4 μm range.

Photodiodes featuring multiple stages of thermoelectric cooling, offering superior sensitivity, are not entirely suitable for developing a compact, lightweight device with low power consumption for integration onto UAVs. This is due to the requirement, in their use, for bulky heat sinks or active cooling systems to dissipate excess heat generated. In contrast, InGaAs photodiodes do not possess these drawbacks; moreover, such photodiodes, meeting the necessary spectroscopic quality standards, are significantly more affordable owing to advancements in telecommunication technologies.

It is pertinent to mention that commercially available diode lasers emitting at a central wavelength of 1.65 μm offer emission powers of up to 100 mW. In contrast, emission powers at a wavelength of 3.24 μm are typically half an order of magnitude lower, albeit at a significantly higher cost. Furthermore, intricate alignment procedures are often unnecessary in the near-infrared range, as photonic elements with fiber input/output of radiation can be employed. Standard quartz optical fibers are utilized up to a wavelength of ~2.3 μm. These considerations underpin the near-infrared range selection as the preferred working range.

The method for remotely sensing the abundance of atmospheric gas species relies on continuous measurement of the absorption line depth. Given that the proposed technique operates effectively for an isolated spectral line, the $R_4$ multiplet of the first overtone of the $2\nu_3$ ($F_2$) band Q-branch at a wavelength of 1651 nm was selected as the working range. The simulation of this spectral line, based on the Voight profile parameters from the HITRAN-2020 database, is illustrated in Figure 2 [28-30]. Examination of the figure reveals that, at atmospheric pressure, the full width at half maximum (FWHM) of the absorption line $R_4$ is approximately $\Delta\nu \approx 0.15$ cm$^{-1}$. This line, devoid of overlap with the absorption lines of $H_2O$, remains observable even under conditions of moderate to high humidity.

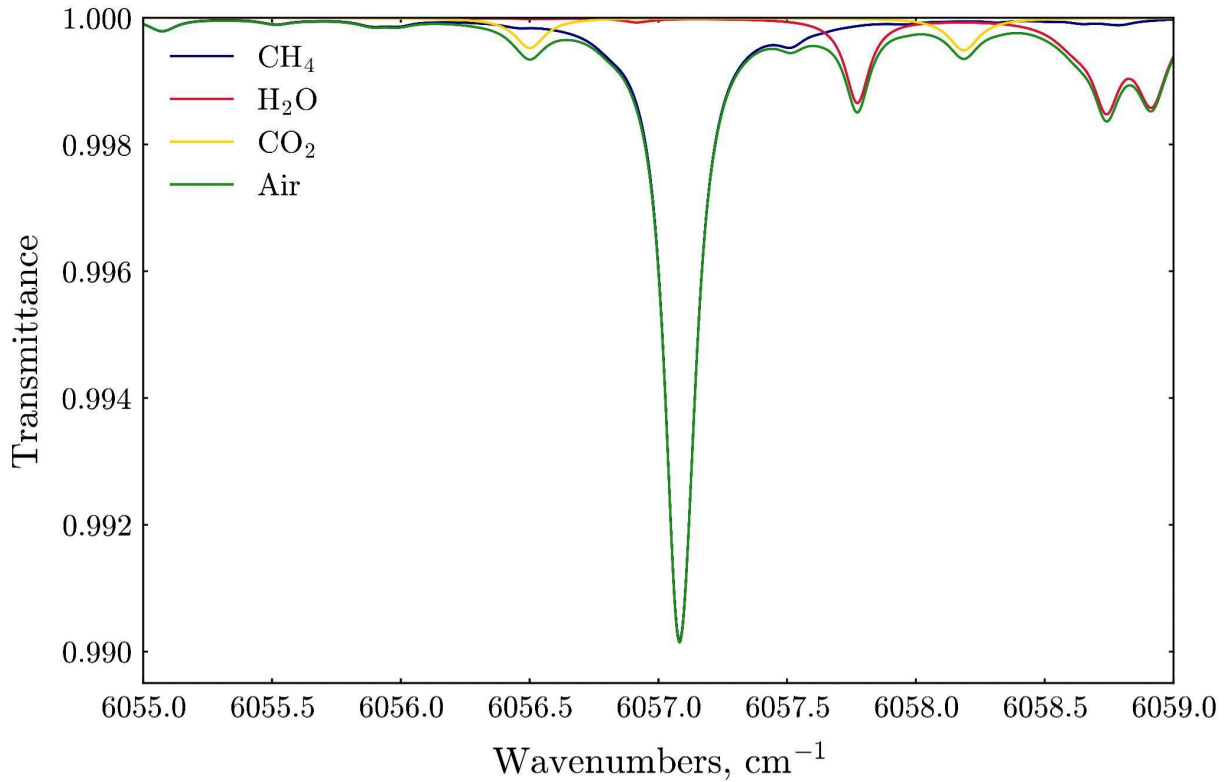

*Fig. 2. Spectral transmittance of atmospheric air gas species within the 6055-6059 cm$^{-1}$ range at 298 K and a distance of 100 m, derived from the HITRAN-2020 database [28-30].*

### 2.2. Data treatment

Modulation spectroscopy offers the potential for better sensitivity than classical tunable diode laser absorption spectroscopy (TDLAS). It presents two primary advantages: firstly, it allows measurement of a difference signal directly proportional to the species concentration in a low-absorbance limit (less than approximately 10% absorbance) [31], and secondly, it shifts the measured signal to a higher frequency region, thus enhancing the signal-to-noise ratio and consequently increasing sensitivity.

This technique encompasses two distinct approaches. Wavelength modulation spectroscopy (WMS) [32,33] involves modulation frequencies much smaller than the width of the absorption spectral line of the target gas, with the signal analyzed at the modulation frequency *f* or its harmonics (*2f, 3f*, etc.). Modulation frequencies ranging from a few kHz to several MHz are commonly utilized. It was shown that the maximum *2f*-signal is obtained for modulation amplitude of the half width at half maximum (HWHM) of the absorption line multiplied by 2.2 [33]. This method has been in use since the early 1970s, employing tunable diode lasers [34]. Conversely, frequency modulation spectroscopy (FMS) employs modulation frequencies comparable to or greater than the width of the spectral feature of interest [35,36]. In this study, the WMS method has been employed and thus will be elaborated upon subsequently.

The WMS method entails high-frequency modulation of the laser injection current, leading to a harmonic variation of the laser emission wavelength over time. This approach mitigates baseline issues, as the signal received by the photodetector exhibits a horizontal baseline due to the modulation form, which minimally impacts the outcome. Moreover, employing a synchronous detector as a receiving device enhances the signal-to-noise ratio by more than an order of magnitude compared to results obtained using direct detection techniques [37].

The accurate operation of this technique necessitates achieving dynamic stabilization of the central wavenumber at a precision level better than $10^{-3}$ cm$^{-1}$. Such a degree of stabilization is attainable by utilizing the wavenumber of the absorption line peak of the gas under examination as a reference. However, in the majority of WMS applications, only laser crystal temperature control employing a Peltier element is applied, which can ensure stabilization accuracy of $10^{-2}$ cm$^{-1}$ [38-41].

The modulation assumes a sinusoidal shape. The signal the synchronous detector receives at the modulation frequency *f* is proportional to the variation in incident laser power, which is contingent on the wavelength. The first harmonic of the detected signal while scanning the absorption line, exhibits a profile resembling the derivative of the absorption line. The *1f*-signal carries the information on the net flux of the received radiation. Determining the distance to the surface scattering the laser radiation is also feasible based on its phase shift. Additionally, the second and third harmonics of the signal at double and triple the modulation frequency are calculated. Without spectral features in the scanned region, the *2f*- and *3f*-signals will register as zero. The signal's second harmonic denotes the intensity of the selected spectral absorption line. Conversely, the third harmonic is beneficial for laser radiation stabilization since, unlike the first harmonic, it lacks an offset, and its point of symmetry aligns with the center of the absorption line. An illustration of the first three harmonic components of the received signal behavior during the scanning of the selected methane absorption spectral line is shown in Figure 3.

The impact of laser power variation corresponding to the modulated injection current can be mitigated through the 2f-signal component analysis, resembling the second derivative of the line profile [42]. This approach diminishes the 1/*f* laser flicker noise and minimizes the measurement sensitivity to thermal fluctuations. These benefits enable absorbance measurements at the $10^{-6}$-$10^{-7}$ level with high-frequency wavelength modulation in the 10 MHz range [43]. However, a drawback of this method is the necessity for calibration due to the challenge of solving the inverse problem.

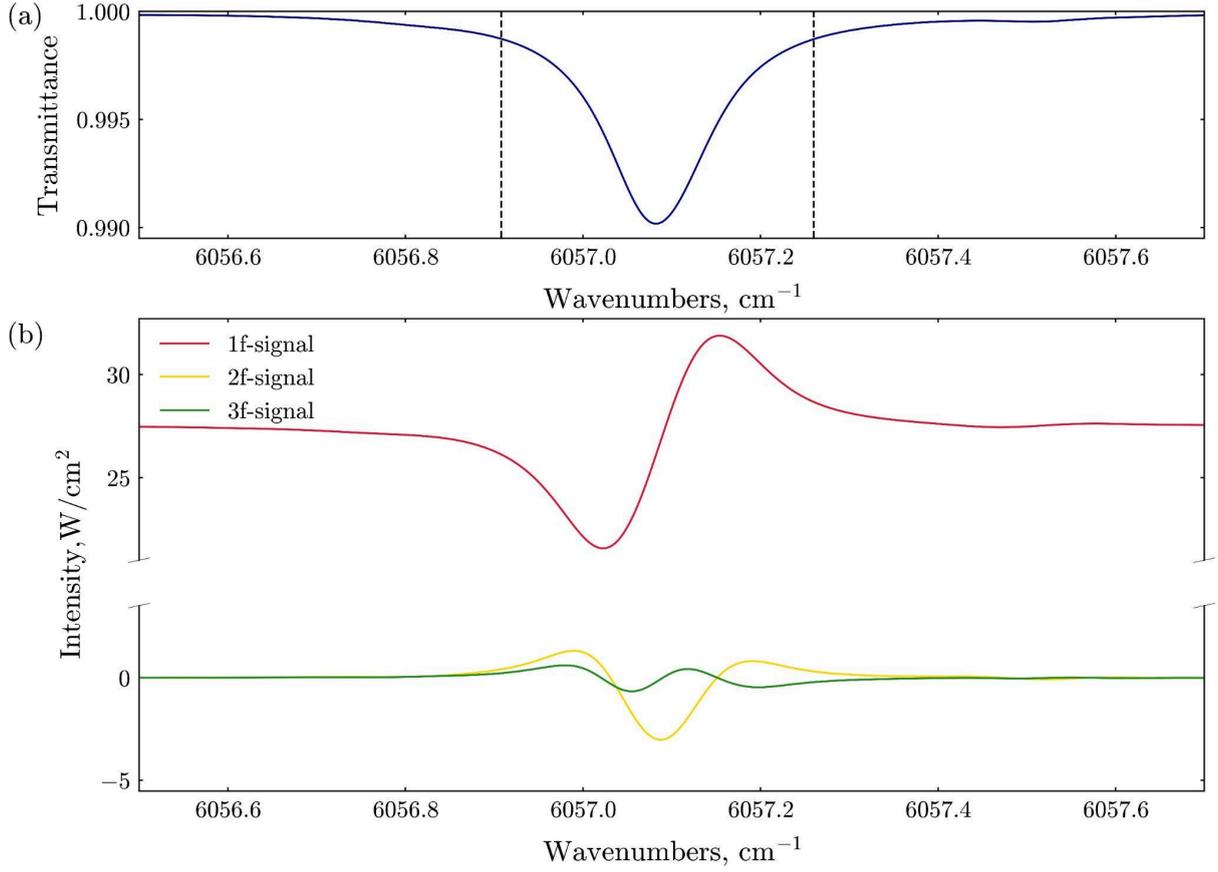

*Fig. 3. (a) Transmittance of methane with column density of 1000 ppm·m in the R4 multiplet of the 2v₃ (F₂) band Q-branch, corresponding to a wavelength of 1651 nm, based on the HITRAN-2020 database [28-30], vertical dashed lines show the optimum scanning range [33]; (b) modeling of the harmonic components of radiation passed through atmospheric air with methane content at a level of 1000 ppm·m at the f, 2f, and 3f frequencies.*

Detailed discussion on the formation principles of radiation scanning within the chosen spectral region according to the WMS with QD of the received signal is presented in our previous paper [25]. Here, we briefly present an analytical model that illustrates the method. The maximum contribution of the intensity of the acquired signal $P$ [W/cm²] at the modulation frequency $f$ will be as follows

$$P(f)_{max} = K \cdot \Delta I, \qquad (1)$$

where $\Delta I$ represents the amplitude of radiation intensity sinusoidal modulation [W/cm²], $K$ denotes the coefficient of radiation loss along the optical path length from the radiation source to the photodetector, which depends on the diameter $D$ of the lens receiving the scattered radiation and the distance $L$ to the scattering surface:

$$K \approx \frac{\pi \frac{D^2}{4}}{\pi L^2} = \left(\frac{D}{2L}\right)^2. \qquad (2)$$

In the synchronous signal detection at the doubled frequency *2f* scenario, when the laser-generated radiation frequency aligns with the spectral absorption line peak position, the signal intensity can be expressed as follows:

$$P(2f) \approx K \cdot I_0 \cdot N \cdot L, \tag{3}$$

here, $I_0$ denotes the constant component of radiation intensity sinusoidal modulation [W/cm$^2$], $N$ – concentration of the analyzed gas [mol/cm$^3$].

Subsequently, combining equations (1) and (3), we derive:

$$\frac{P(2f)}{P(f)} \approx \frac{I_0}{\Delta I} \cdot N \cdot L, \tag{4}$$

hence, it is possible to determine the detected gas concentration from the signals at the modulation frequency $f$ and the doubled frequency $2f$. It is noteworthy that the signal at twice the frequency $2f$ will be inversely proportional to the distance to the scattering surface $L$, while the signal at the modulation frequency $f$ will be inversely proportional to $L^2$, and their ratio will be proportional to $L$.

The determination of the distance to the radiation scattering surface, as previously discussed, could be derived from the phase shift of the detected signal:

$$H = \frac{L}{2} = \frac{1}{2}\frac{c}{f}\frac{\Delta\varphi}{2\pi}, \tag{5}$$

where $c$ denotes the speed of light. However, in practice, this procedure may necessitate calibration since the phase shift $\Delta\varphi$ encompasses the delay associated with radiation propagation and the effects of analog electronics, namely the trans-impedance amplifier (TIA) and bandpass filter, on the signal. Such an effect could be mitigated through coherent detection, as mentioned in our previous work regarding Frequency Modulated Continuous Wave lidar [44].

## 3. Instrument Design

UAV payload limitations defined the shape of our instrument. Specifically, the payload mass should not surpass 5 kg to facilitate extended flights on a single UAV battery charge. Additionally, the dimensions of the UAV impose constraints on the maximum payload size.

### *3.1. Optics*

The instrument body consists of two units: the control electronics and reference channel unit, as well as the analytical channel unit, with the flexibility to adjust their relative positions as needed.

The control electronics and reference channel unit house the main control board, laser driver board, diode laser with fiber output, external optical isolator, fiber beam splitter, fiber collimator, N-BK7 glass reference cell, photodetector (PD), batteries, and cooler. For the radiation source, we opted for the DFB diode laser by LasersCom at a central wavelength of 1.65 μm, an output power of approximately 8 mW, and a spectral width of 500 kHz.

Laser radiation is coupled into a fiber beam splitter with a 97/3 ratio, directing most of the radiation to the analytical channel and a smaller portion to the reference channel, with around 10% of the power lost at the fiber couplers. In the reference channel, the collimated radiation traverses a 46.5 mm long N-BK7 glass cell containing a $CH_4$ and $N_2$ mixture before reaching the PD. The photodetector is an InGaAs photodiode from ThorLabs (FGA10), connected to a TIA board designed to handle the *3f*-signal. The stabilization algorithm is implemented based on the methane absorption line peak, coinciding with the symmetry point of the reference channel *3f*-signal, as depicted in Figure 3.

The analytical channel of the device consists of a plano-convex N-BK7 condenser lens, ThorLabs F280APC-1550 long-focus fiber collimator, an aluminum alloy cone replicating the geometry of the convergent beam of the received radiation, and an InGaAs photodiode ThorLabs FGA21 with a TIA board enclosed in a metal case to isolate from external noise. The principal optical diagram of the GIMLI prototype is illustrated in Figure 5.

The decision to incorporate the plano-convex N-BK7 condenser lens, with a diameter of 100 mm and an effective focal length of 136.2 mm, necessitated the adoption of a misalignment scheme between the laser fiber collimator and the receiving optics. This adjustment, illustrated in the optical layout of the analytical channel unit shown in Figure 4, was required due to the challenges associated with placing the optical elements on a single axis. The application of this configuration presented a challenge regarding the minimum operating distance to the scattering surface, as it was optimized for a distance of 50 m.

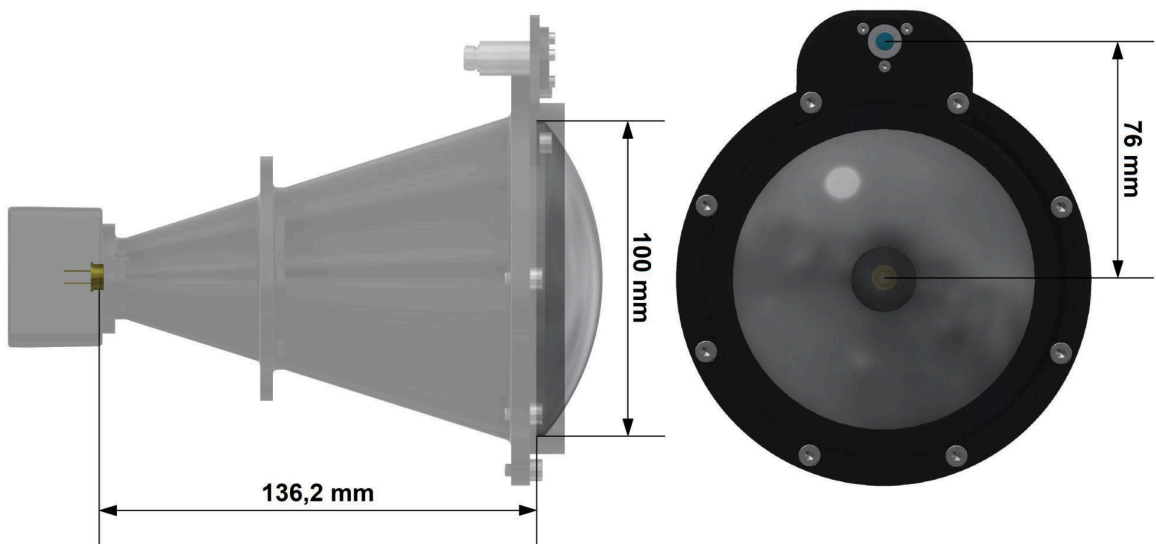

*Fig. 4. Optical scheme illustrating the analytical channel unit. The sensitive area of the photodiode is located at the focal point of the lens collimating the received radiation.*

The units are linked by optical fiber, signal cable, and power cable, with lengths allowing for various configurations of their relative positions. If the height of the UAV support legs is insufficient, the units could be arranged as depicted in Figure 1b.

### *3.2. Electronics*

For the GIMLI prototype, we developed the control electronics implemented across several circuit boards: digital control board, laser driver, TIA for photodiodes, battery charger, and power supply. The block diagram of the electric circuits is illustrated in Figure 5.

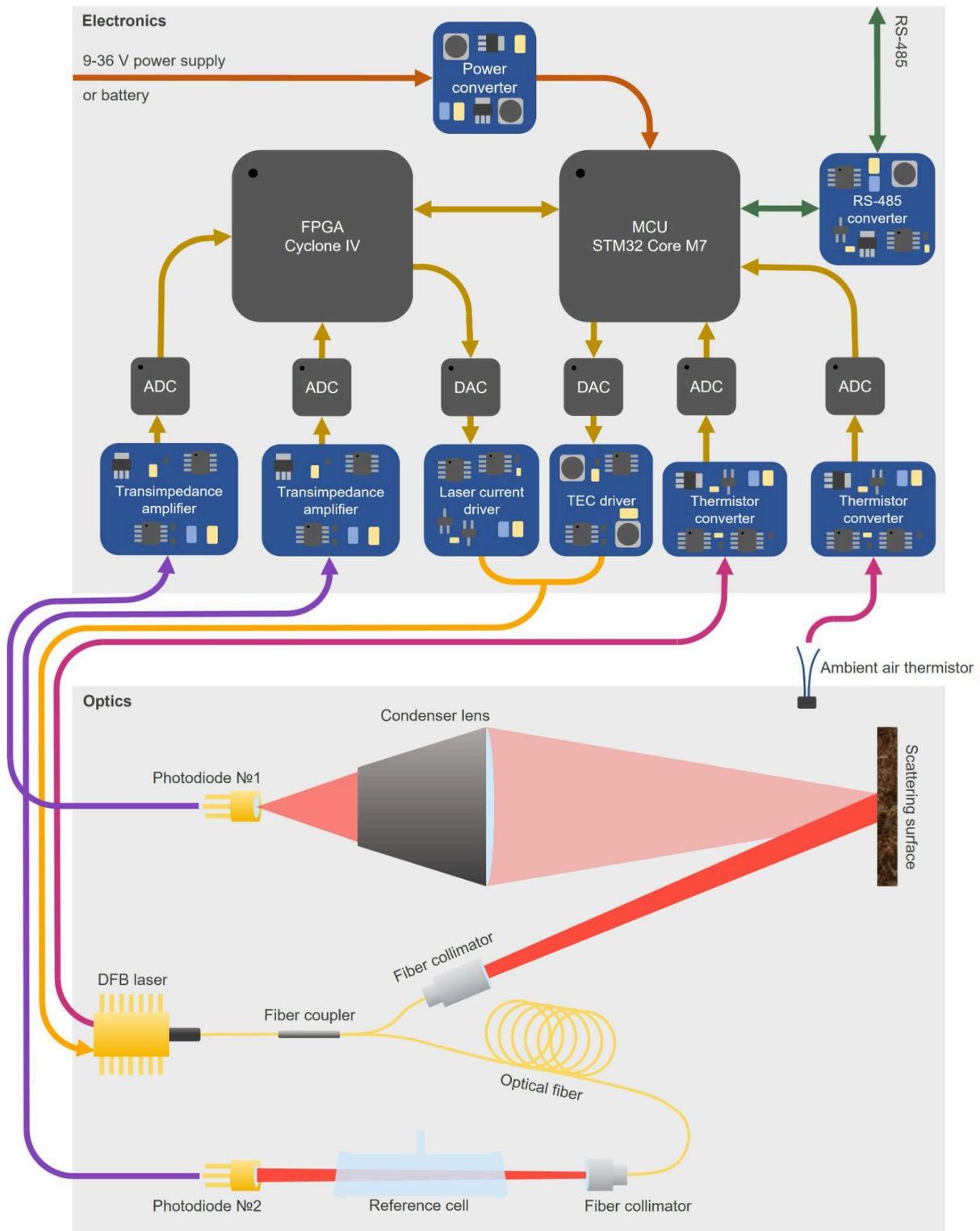

*Fig. 5. Block diagram of the electric circuits and principal optical diagram of the GIMLI prototype.*

The analog module of the device prototype is designed to adjust the diode laser injection current and temperature, as well as to amplify the received signal. The digital board includes four ADCs and two DACs, a programmable microcontroller (MCU), and a field-programmable gate array (FPGA).

A rapid 16-bit DAC controls the laser pump current, and another 16-bit DAC adjusts the voltage on the Peltier element integrated into the laser housing. Two 12-bit ADCs capture the pre-amplified and filtered signals from the analytical and reference channels' photodiodes, while the 16-bit ADC registers the voltage from the built-in thermistor to calculate the current temperature of the laser crystal and another 12-bit ADC registers the B57861-S NTC thermistor voltage to record the ambient air temperature.

The chosen Cyclone IV FPGA, in conjunction with the employed data processing algorithm, enables the output of pre-processed data at a modulation frequency of $f$ = 26 kHz and a sinusoidal modulation period of 192 points averaged over 1024 times. However, the repetition frequency of a complete cycle of spectral measurements is constrained by the maximum operating rate of the STM32 MCU with ARM Cortex-M7 core, approximately 18.5 Hz. The detailed characteristics of the developed GIMLI prototype are presented in Table 1.

*Table 1. Specification of the GIMLI prototype.*

| | |
|---|---|
| Sensitivity | 50 m → 28 ppm·m<br>75 m → 42 ppm·m<br>100 m → 56 ppm·m |
| Operating distance (min/max) | 40/120 m |
| Mass | ~4 kg |
| Dimensions (WHD) | 275×255×175 or 275×190×400 |
| Aperture of receiving optics | 100 mm |
| Radiation wavelength | 1651 nm |
| Laser power | ~8 mW |
| Sampling rate | ~19 Hz |
| Power consumption (standard/peak) | 12/35 W |
| Power supply | built-in battery(7.4 V)/<br>on-board power supply(9-36 V) |

## 4. Results

During the field tests of the GIMLI prototype, its performance was evaluated under various experimental setups, including onboard a UAV and in a horizontal configuration (where the external optical unit of the device was mounted on a tripod, capturing scattered signals from a vertically positioned screen). These tests were conducted across all seasons typical for middle latitudes, in the temperature range of about 40℃, in both dry air and light precipitation and different surfaces scattering radiation.

*4.1. Experimental validation of analytical dependencies*

We examined the behavior of different signal characteristics based on the distance to the scattering surface, measuring background methane at an ambient temperature of 26℃. As previously noted, the misalignment of the laser radiation fiber collimator and the optical axis of the analytical channel detecting system necessitated a minimum working distance between the gas analyzer and the scattering surface of approximately 40 m, due to the optical system's adjustment for a distance of 50 m.

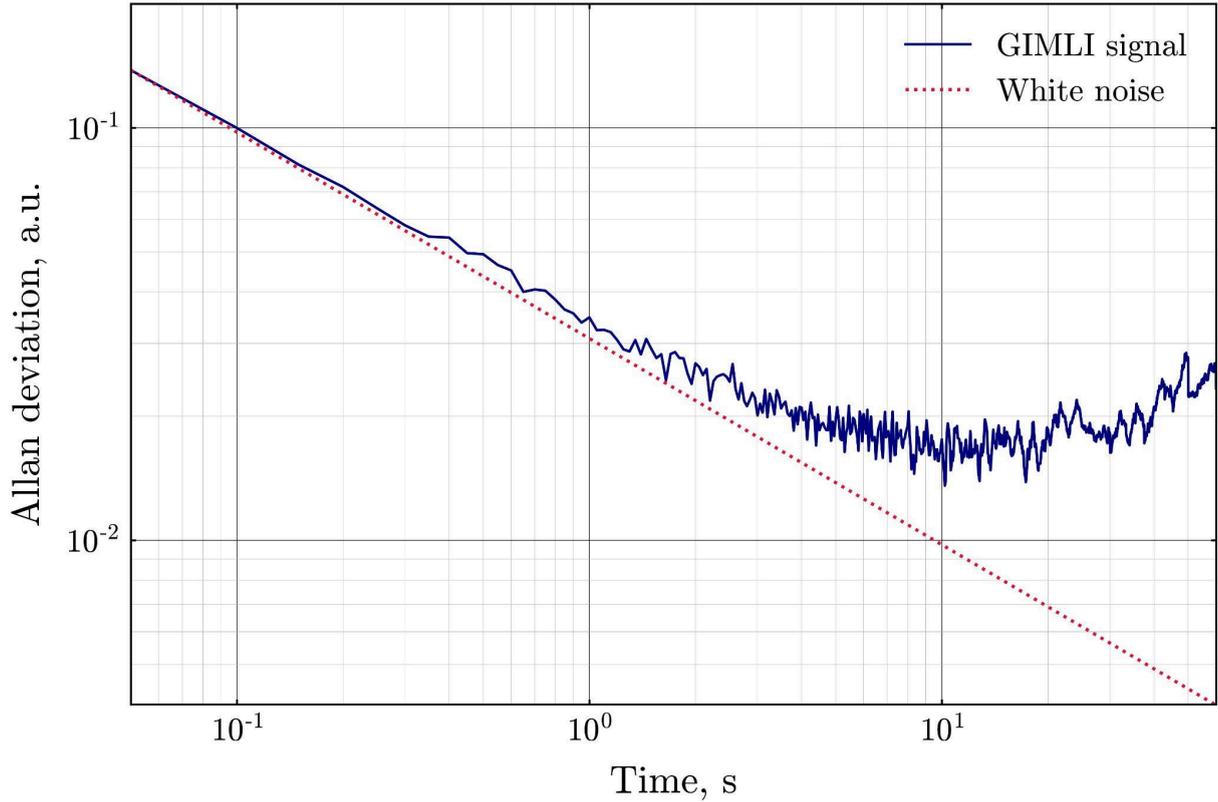

*Fig. 6. Comparison between the Allan deviation for the signal acquired at a distance of 50 m from the scattering surface (blue curve) and the corresponding dependence for white noise (red dots).*

The primary mode of instrument operation involves measuring the background methane integral concentration in the atmospheric air while onboard a UAV moving at an altitude of approximately 50 m with a speed ranging from 10 to 15 m/s. Under these conditions, the potential signal averaging times, ensuring spatial resolution finer than tens of meters, will not exceed a few seconds. Figure 6 illustrates the dependence of the Allan deviation [45] for the signal acquired at a distance of 50 m from the scattering surface on the accumulation time, alongside its comparison with the corresponding dependency for white noise, presented on a double logarithmic scale. Signal stability on time scales of less than tens of seconds, caused by the open path analytical channel optical system, is deemed optimal for the expected device applications.

The chief attribute of the sine form signal received, influencing subsequent analyses, is its peak-to-peak amplitude, depicted in Figure 7. As per equation (2), the peak-to-peak amplitude of the scattered radiation signal diminishes in inverse proportion to the square of the distance to the scattering surface. Figure 7 displays the experimental relationship between the detected signal peak-to-peak amplitude and the distance to the scattering surface, alongside its approximation using a least squares method-derived function of the form $1/x^2$.

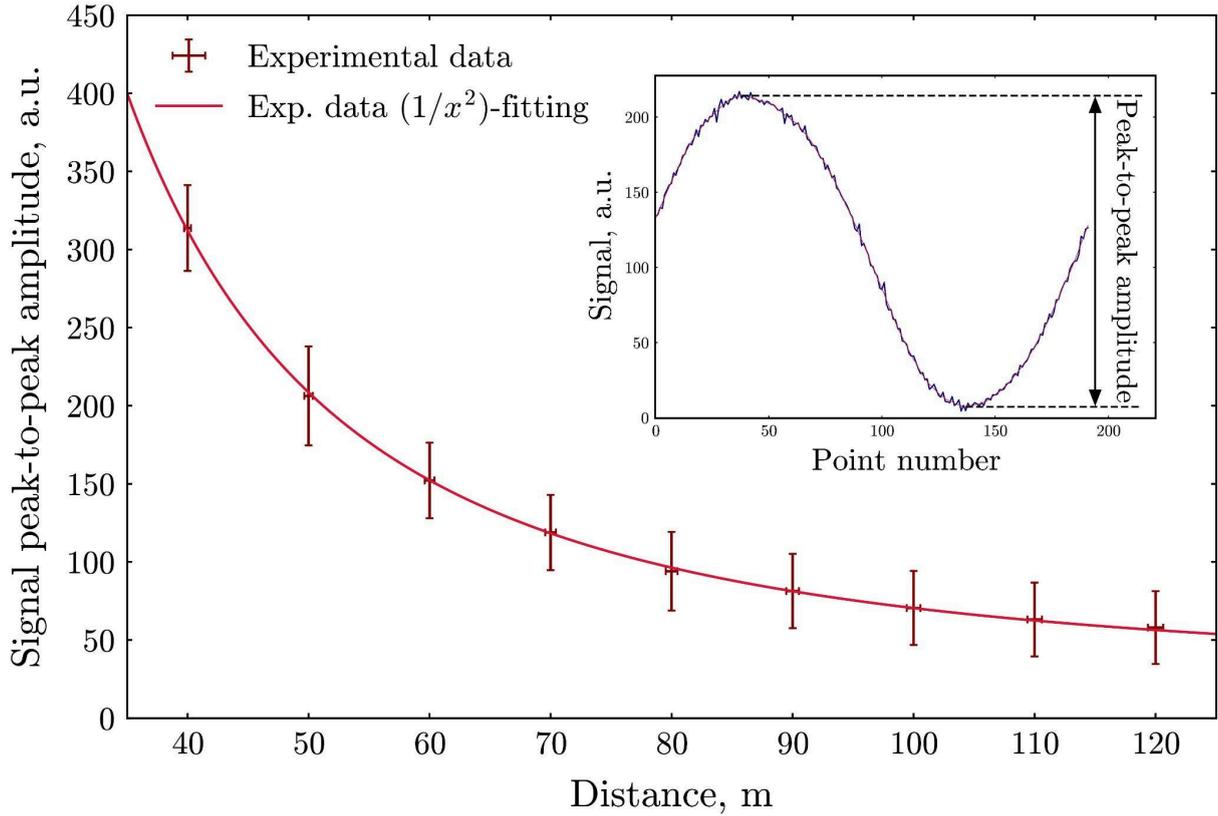

*Fig. 7. Relationship between the signal peak-to-peak amplitude and the distance to the scattering surface (red crosses), along with its approximation using a function of the form $1/x^2$ (red curve) and recorded signal (in the upper right corner).*

The recorded signal, depicted in Figure 7, comprises *n* harmonic components spanning from *f* to *nf* frequencies. Utilizing the QD method, it becomes possible to select the signal harmonics essential for subsequent analysis, as demonstrated in [25]. The *1f*-signal, representing the total intensity of the received radiation, as per equation (1), exhibits a dependence on the distance to the scattering surface similar to the peak-to-peak amplitude of the recorded signal, as depicted in Figure 8. Similarly, the *2f*-signal, indicative of the selected spectral line absorption intensity according to equation (3), varies with the distance to the scattering surface as $1/x$, as also illustrated in Figure 8.

Hence, the ratio of the *2f*-signal to the *1f*-signal, as inferred from (4), should exhibit linearity concerning the distance to the laser-scattering surface. Figure 9 illustrates this relationship, depicting the obtained ratio for the GIMLI device along with its linear approximation.

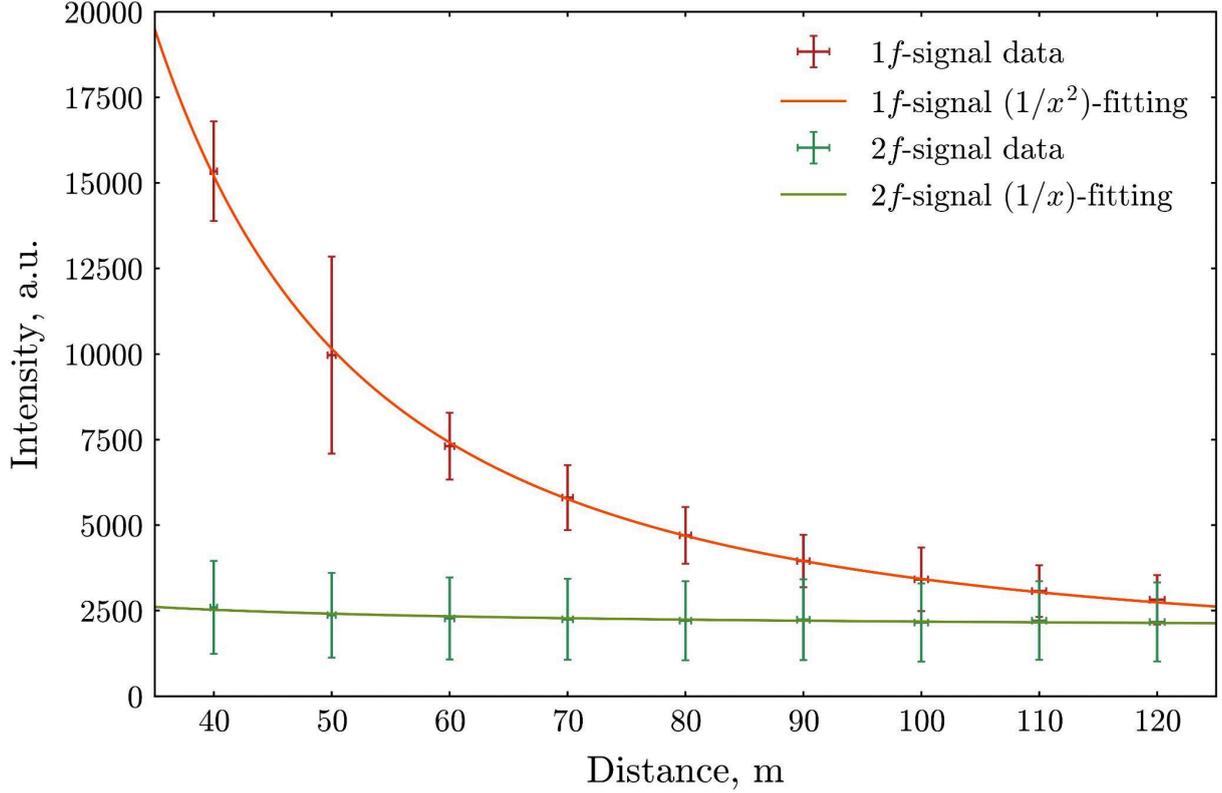

*Fig. 8. Relationships between the 1f-signal (red crosses) and 2f-signal (green crosses) concerning the distance to the scattering surface, along with their respective approximations using functions of the form 1/x² (red curve) and 1/x (green curve), respectively.*

### 4.2. The GIMLI prototype calibration

To convert the GIMLI prototype methane abundance measurements into ppm·m units, the instrument underwent calibration through independent measurements from the LI-COR LI-7810 precision gas analyzer. The LI-7810 analyzer capable of determining $CH_4$ concentration with a precision of 2 ppb according to the manufacturer's specifications, served as the reference. The experimental setup involved the GIMLI prototype mounting on a tripod, and a vertical screen set at varying distances from the instrument, ranging from 40 to 120 m in 10 m intervals. The ambient air temperature was approximately 26℃ throughout the experiment.

The LI-7810 gas analyzer features continuous air pumping through the analytical cell to ensure sample refreshment during measurements. During our experiment, the operator of the LI-7810 device repeatedly moved back and forth along the optical path of the GIMLI prototype's laser radiation. The LI-7810 readings $N_{LI-COR}$ measured in ppm, were updated at a frequency $f_{LI-COR}$ = 1 Hz. Thus, given the operator's average speed along the optical path $v_{op}$, methane abundance in the air, expressed in ppm·m for each distance $L$, was obtained using the formula:

$$N_{ppm \cdot m} = \sum_{l=0}^{L} \left[ \frac{N_{LI-COR} \cdot v_{op}}{f_{LI-COR}} \right]. \tag{6}$$

Subsequently, the obtained values were averaged based on the number of passes along the optical path. The relationship between the integral methane concentrations, calculated using LI-7810 gas analyzer data, and the measurement distance *L* from 40 to 120 m at 10 m intervals, is depicted in Figure 10.

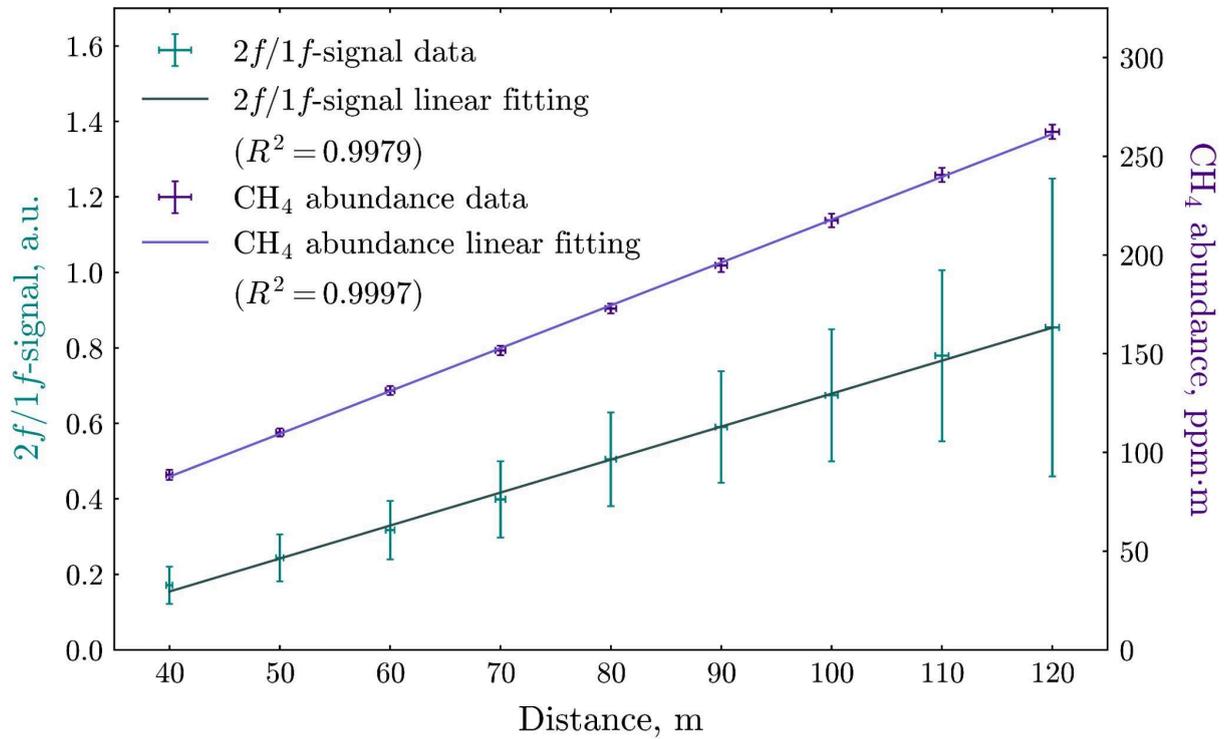

*Fig. 9. Relationship between the 2f- to 1f-signal ratio and the distance to the scattering screen: experimentally acquired values with measurement errors (green crosses) and linear approximation of the experimental data (green line). Dependency of integral methane concentrations calculated from LI-7810 data on the working distance: experimentally acquired values with measurement errors (purple crosses) and linear approximation of the experimental data (purple line).*

As anticipated, the relationship exhibited linearity with distance due to the effective mixing of background methane in the atmospheric air. The primary factor contributing to the error in determining the value of $N_{ppm \cdot m}$ was inaccuracies in measuring the distance between the GIMLI prototype and the screen using a construction laser rangefinder. Data collected by the GIMLI prototype was recorded at identical time intervals to those of the LI-7810 gas analyzer. Subsequently, through processing the data acquired by the GIMLI prototype, the correlation between the *2f-* to *1f-*signal ratio and the distance to the scattering screen was established, as depicted in Figure 10. Notably, this correlation also conforms to a linear approximation, aligning with the isotropic distribution of methane abundance in the atmospheric air along the optical path. Subsequently, the relationship between the *2f-* to *1f-*signal ratio and the value of $N_{ppm \cdot m}$ for the respective distances was plotted (Figure 10).

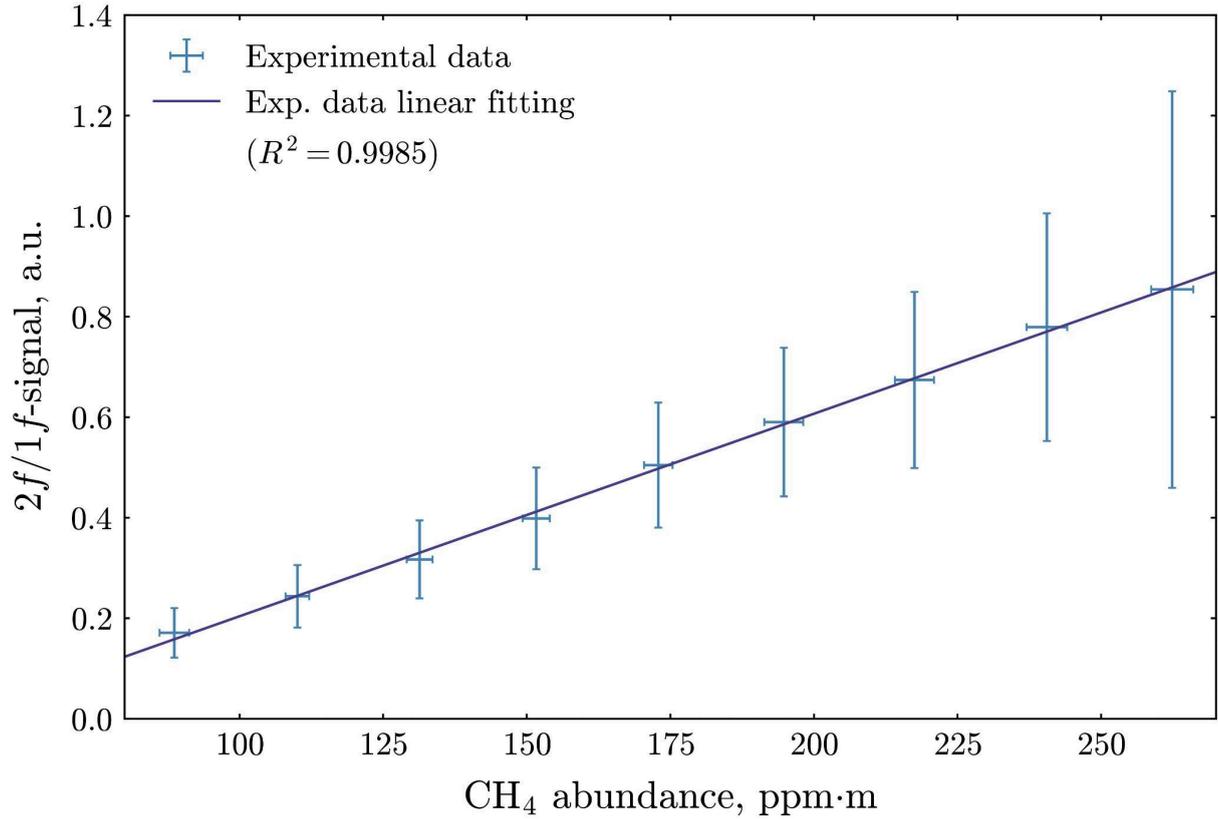

*Fig. 10. Relationship between the 2f- to 1f-signal ratio and integral methane concentrations derived from LI-7810 gas analyzer data at various distances. Experimentally obtained values with measurement errors (green crosses), the linear approximation of this relationship (purple line).*

The coefficients derived from the linear approximation of the *2f-* to *1f-*signal ratio values' dependence on the $N_{ppm \cdot m}$ values were adopted as calibration coefficients for future methane abundance measurements in atmospheric air using the GIMLI prototype. Along with the calibration results, the sensitivity of the device with respect to the distance to the scattering surface was obtained. The data collected over 10-second intervals at each distance were utilized to ascertain the sensitivity at various distances from the radiation-scattering surface. Employing the mean *M* values and standard deviation derived from the calculated *2f-* to *1f-*signal ratios, the sensitivity at each distance was determined using the following equation:

$$D_{GIMLI} = N_{ppm \cdot m} \cdot \frac{STD\left[\frac{2f-signal}{1f-signal}\right]}{M\left[\frac{2f-signal}{1f-signal}\right]}, \quad (7)$$

where $N_{ppm \cdot m}$ denotes the methane concentration in ppm·m units according to LI-7810 gas analyzer measurements.

The upper limit of the working distances is approximately 120 m determined by a signal-to-noise ratio. The standard deviation-to-mean ratio of the *2f-* to *1f-*signal ratio remains consistent across the entire range of operational distances. Consequently, the instrument's sensitivity is expected to vary linearly with the distance to the scattering surface, given the linear dependency of $N_{ppm \cdot m}$ on distance due to the effective mixing of background methane in atmospheric air.

A comparison of the sensitivity of the presented gas analyzer concerning the distance with the claimed sensitivities of commercially available methane gas analyzers suitable for installation on small-capacity UAV platforms is depicted in Figure 11.

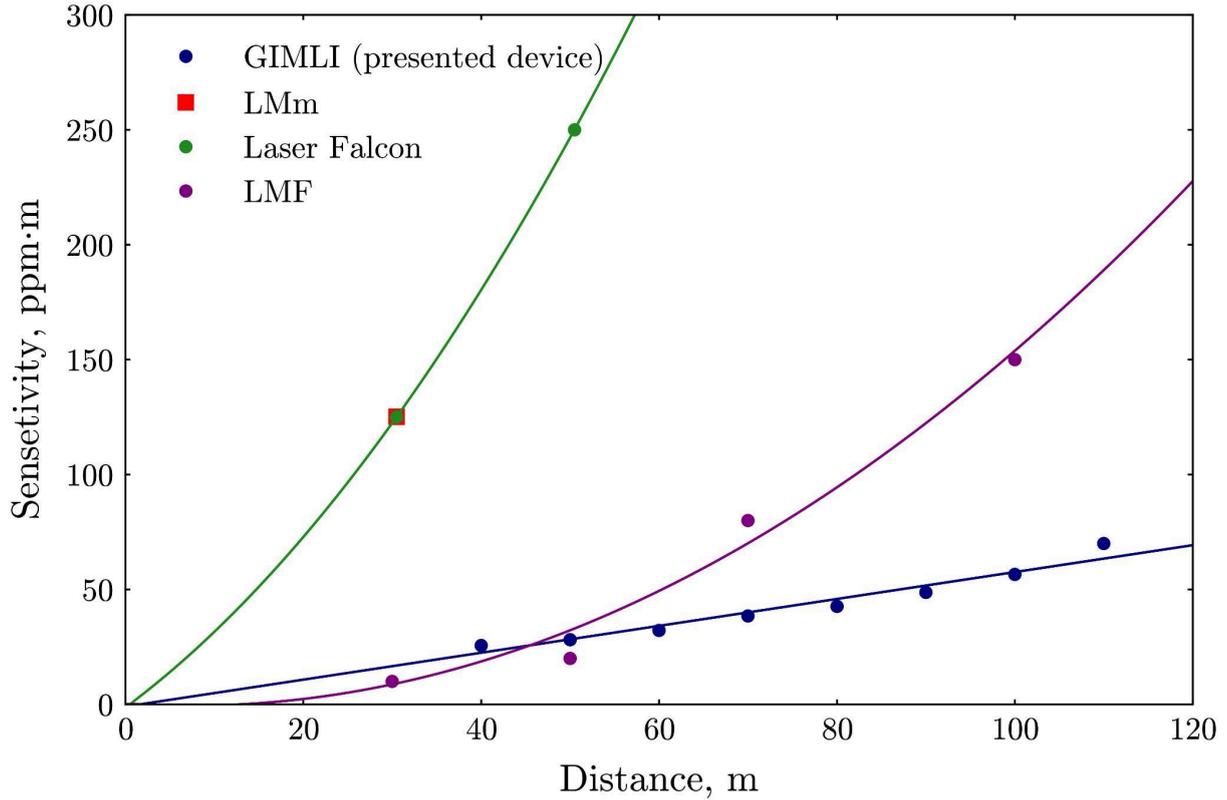

*Fig. 11. Comparison of the sensitivity of the presented gas analyzer concerning the distance with the claimed sensitivities of commercially available methane gas analyzers suitable for installation on small-capacity UAV platforms – a lower value for the selected distance corresponds to better sensitivity.*

The implementation of the proposed WMS method with QD has validated equation (5), provided that the phase shift due to the phase response of the analytical channel TIA is taken into account, including the proportional coefficient and offset. Hence, calibration for distance determination becomes imperative. The calibration results are shown in Figure 12. The ability to autonomously determine the distance to the underlying surface is essential for accurate analysis of selected gas component concentration in ppm·m units, particularly in instances of anomalies.

To summarize, the developed GIMLI prototype can effectively perform remote sensing of methane abundance in atmospheric air, enabling the measurement of methane's integral concentration in ppm·m units with a sensitivity level of less than 30 ppm·m on the typical altitude of 50 m (which is equal to 300 ppb or ~13% of atmospheric air methane abundance) and altitude working range of 40-120 m, while independently determining the distance to the scattering surface.

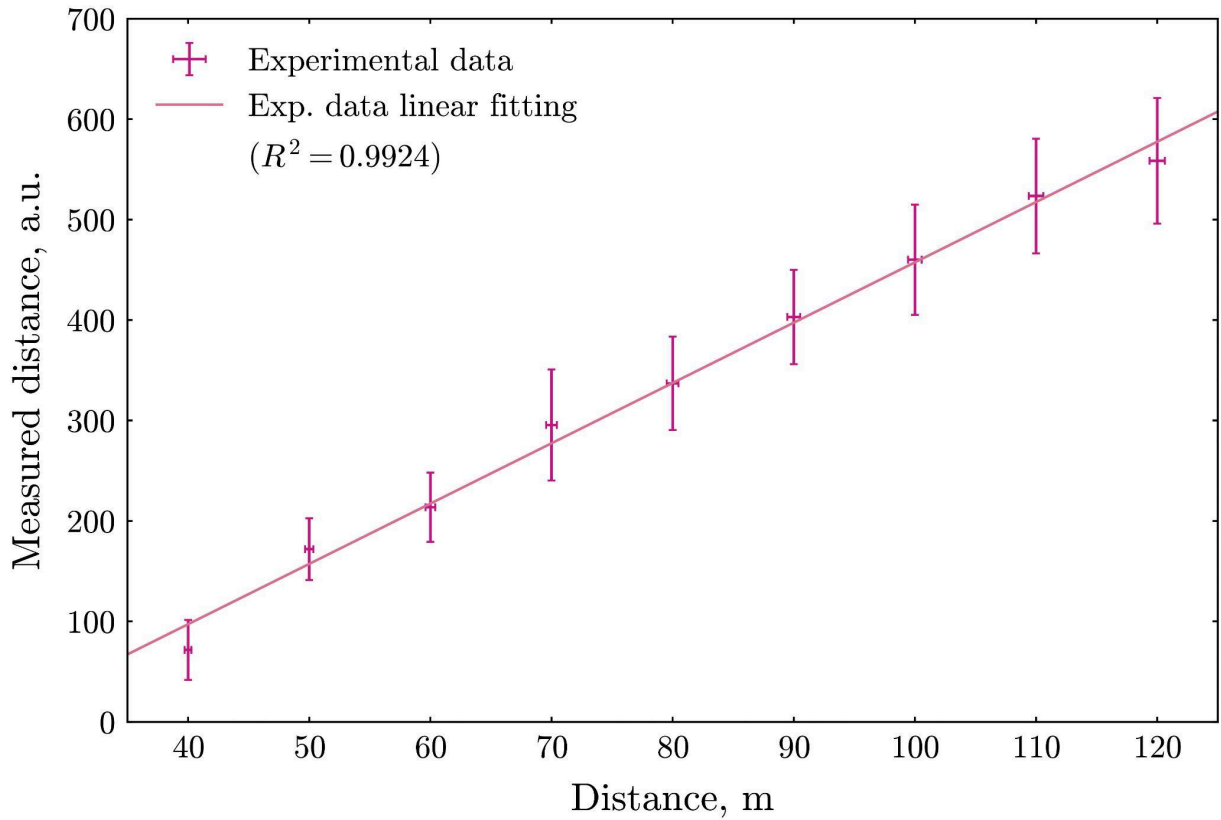

*Fig. 12. Results of the calibration for distance determination technique.*

## 5. Conclusion

In this paper, we experimentally validated the performance of the proposed methane sensor. The instrument can monitor methane onboard UAVs in harsh conditions while being compact. The observed correlation between various signal attributes and the distance to the surface, where laser radiation scatters, aligns consistently with analytical expectations.

The dimensions, mass, and power consumption of the developed device conform to the payload requirements of lightweight UAVs, thereby facilitating cost-effective and streamlined monitoring of gas pipeline leaks, assessment of air quality near hazardous industries, landfills, and other anthropogenic sources of methane emissions, as well as in regions characterized by natural methane emissions such as Arctic regions and swampy terrains.

The GIMLI prototype exhibits sensitivity comparable to existing compact laser gas analyzers for methane monitoring aboard UAVs. Notably, it surpasses competitors at altitudes exceeding 50 meters, deemed safer for pilots, with the help of laser radiation stabilization and a linear sensitivity from distance because of the chosen method. As ongoing research in our group shows, prospects for enhancing device sensitivity by integrating more powerful laser sources and larger diameter optics hold promise.

The deployment of the developed gas analyzer for remote methane monitoring in both natural and anthropogenic emission areas promises to advance existing methane emission measurement methodologies, crucial for quantifying greenhouse gas emissions.

## Declarations

**Funding** Partial financial support was provided by the Russian Foundation for Basic Research under Grant Agreement № 18-29-24204.

**Conflict of interest** The authors declare no conflict of interest.

**Author contributions** Conceptualization: MS; Methodology: VM, MS; Formal analysis and data curation: VM; Investigation: VM, VK, MS; Software: VM, VK; Validation: VM, GS; Writing – original draft preparation: VM; Visualisation: VM; Writing – review and editing: MS, AR; Funding acquisition: AR; Supervision: MS, VM. All authors read and approved the final manuscript.